\documentclass[doublecol]{epl2}
\usepackage{graphicx}
\usepackage[normalem]{ulem}
\usepackage{amsmath}
\usepackage{upgreek}
\usepackage{float}
\usepackage{units}
\usepackage{amssymb}
\usepackage[normalem]{ulem}

\renewcommand{\paragraph}[1]{{\it #1.---}}

\let\section\paragraph

\title{Stick-Slip Contact Line Motion on Kelvin-Voigt Model Substrates}

\author{Dominic Mokbel\inst{1} \and Sebastian Aland\inst{1,2} \and Stefan Karpitschka\inst{3}}
\shortauthor{D. Mokbel \etal}
\institute{%
    \inst{1} HTW Dresden, Friedrich-List-Platz 1, 01069 Dresden, Germany\\
    \inst{2} TU Bergakademie Freiberg, Akademiestrasse 6, 09599 Freiberg, Germany\\
    \inst{3} Max Planck Intitute for Dyanmics and Self-Organization (MPIDS), 37077 G\"ottingen, Germany
}

\date{\today}%

\abstract{%
The capillary traction of a liquid contact line causes highly localized deformations in soft solids, tremendously slowing down wetting and dewetting dynamics by viscoelastic braking. Enforcing nonetheless large velocities leads to the so-called stick-slip instability, during which the contact line periodically depins from its own wetting ridge.
The mechanism of this periodic motion and, especially, the role of the dynamics in the fluid have remained elusive, partly because a theoretical description of the unsteady soft wetting problem is not available so far.
Here we present the first numerical simulations of the full unsteady soft wetting problem, with a full coupling between the liquid and the solid dynamics.
We observe three regimes of soft wetting dynamics: steady viscoelastic braking at slow speeds, stick-slip motion at intermediate speeds, followed by a region of viscoelastic braking where stick-slip is suppressed by liquid damping, which ultimately  gives way to classical wetting dynamics, dominated by liquid dissipation.
}%

\begin{document}

\maketitle


\section{Introduction}%
The capillary interaction of liquids with soft solids is a ubiquitous situation in natural or technological settings~\cite{Andreotti:ARFM2020,Bense:2019,Chen:COCIS2018,Style:ARCMP2017}, ranging from droplets that interact with epithelia, for instance in the human eye~\cite{Holly:EER1971}, or epithelial cells governed by capillary physics~\cite{PerezGonzalez:NP2018}, to ink-jet printing on flexible materials~\cite{Wijshoff:COCIS2018}.
The capillary tractions, exerted by the liquid onto their soft support, cause strong deformations if the substrate is soft or the considered length scale is sufficiently small~\cite{Style:PRL2013,Zhao:NL2021}.
The typical scale below which capillarity deforms solids is given by the elastocapillary length, $\ell = \gamma/G_0$, the ratio of surface tension $\gamma$ and static shear modulus $G_0$.
At three-phase contact lines, the length scale of the exerted traction lies in the molecular domain, deforming the solid into a sharp-tipped wetting ridge~\cite{Park:NC2014}.
As a liquid spreads over a soft surface, the traction moves relative to the material points of the substrate.
The necessary rearrangement of the solid deformation leads to strong viscoelastic dissipation which counteracts the motion, a phenomenon called viscoelastic braking~\cite{Carre:N1996,Long:L1996,Karpitschka:NC2015,Zhao:PNAS2018,Dervaux:SM2020,Henkel:SM2021,Coux:PNAS2020,Leong:PF2020,SmithMannschott:PRL2021}.
At small speeds, the motion remains steady~\cite{Carre:N1996,Long:L1996}, whereas at large speeds, unsteady motion, frequently termed stick-slip, has been observed~\cite{Pu:L2008,Kajiya:SM2013,Karpitschka:NC2015,Park:SM2017,Gorcum:PRL2018,Gorcum:SM2020}.
In this mode, the contact line velocity and the apparent contact angle undergo strong, periodic oscillations.
On paraffin gels, Kajiya et al. observed stick-slip motion only in an intermediate velocity range, returning to continuous motion if the speed was increased even further~\cite{Kajiya:SM2013}.

\begin{figure*}\begin{center}%
	\includegraphics[scale=0.65]{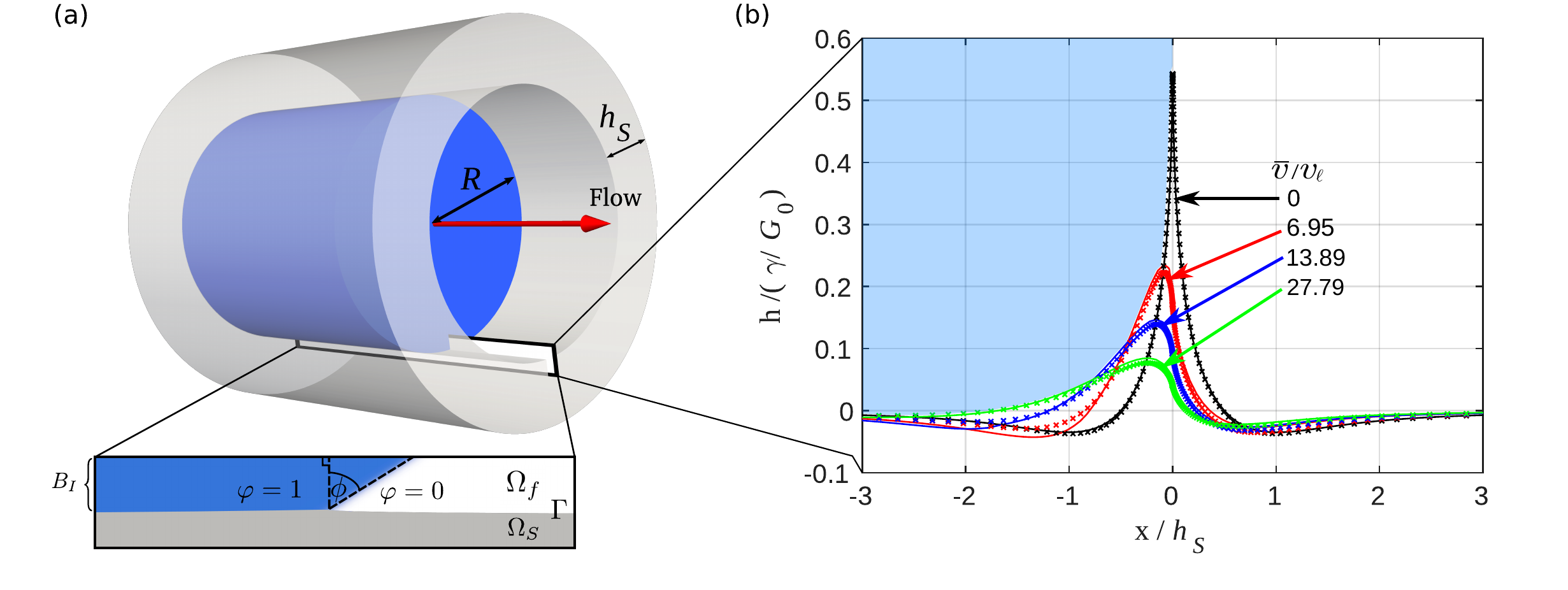}
 \caption{(a)~Dynamical wetting of a cylindrical cavity (radius $R$) with a soft viscoelastic wall (grey, thickness $h_s$) by a two-phase fluid (blue \& transparent). The contact line speed is controlled by the flux boundary condition on the rear end of the cavity. Inset: definition of the liquid-liquid interface rotation $\phi = \theta-\theta_{eq}$ relative to the equilibrium angle $\theta_{eq} = 90^{\circ}$. (b)~Quasi-stationary wetting ridge on the cavity wall for different velocities, comparison between FEM simulations (symbols) and the analytical model for constant $v$ (lines). The liquid interface is aligned at $x=0$, the blue region indicates the advancing liquid phase.}%
	\label{fig:simulation_profiles}%
\end{center}\end{figure*}%

The origin of this stick-slip motion remains debated in literature.
It is clear that the pinning and depinning is not associated with permanent surface features, but rather with the dynamics of the wetting ridge itself: the solid deformation cannot follow the fast contact line motion of the depinned (slip) phase of a stick-slip cycle~\cite{Gorcum:PRL2018,Gorcum:SM2020}.
Unclear, however, are the conditions upon which a contact line may escape from its ridge, thus eliminating the viscoelastic braking force.
The depinning of a contact line from a sharp-tipped feature on a surface is governed by the Gibbs inequality~\cite{Dyson:PF1988,Gorcum:PRL2018}.
Van Gorcum et al. \cite{Gorcum:PRL2018} postulated a dynamical solid surface tension, which would alter the local force balance and thus allow the contact line to slide down the slope of the ridge.
Still, the physico-chemical origin of such dynamic solid surface tension remains elusive.
Roche et al.~\cite{Dervaux:SM2020} postulated the existence of a point force due to bulk viscoelasticity, but the shear-thinning nature of typical soft polymeric materials would prevent such a singularity at the strain rates encountered in soft wetting~\cite{Karpitschka:PNAS2018,Gorcum:SM2020}.
Unclear as well is the role of the fluid phase during the cyclic motion, mainly because a comprehensive multi-physics model for the unsteady soft wetting problem is not available to date.

Here we present the first fully unsteady numerical simulations of dynamical soft wetting, fully accounting for liquid and solid mechanics, and for the capillarity of the interfaces, by which we reveal the life cycle of stick-slip motion.
We derive phase diagrams of steady and unsteady contact line motion by tuning parameters over large ranges, recovering stick-slip behavior at intermediate speeds. At small and large speeds, we observe steady motion, in quantitative agreement with an analytical model.

\section{Setup}%
Figure~\ref{fig:simulation_profiles}~(a) shows the geometric setup of the numerical simulations. A hollow cylinder (undeformed inner radius $R$), made of a soft viscoelastic material (gray, thickness $h_s\ll R$), with a fixed (rigid) outer surface, is partially filled with a liquid (blue) and an ambient fluid phase (transparent).
To keep physics conceivable, we use a minimal model and apply the Stokes limit and identical viscosities for fluid and ambient, and assume constant and equal surface tensions $\gamma=\gamma_s$ on all interfaces (liquid-ambient, solid-liquid, and solid-ambient).
The inner surface of the soft viscoelastic cylinder wall is deformed into a wetting ridge due to the capillary action of the liquid meniscus (cf. panel~(b)).
We use an incompressible Kelvin-Voigt constitutive model, characterized by a frequency dependent complex modulus $G^* = G_0 + i\,\eta_s\,\omega$, with static shear modulus $G_0$ and effective substrate viscosity $\eta_s$, obtaining an elastocapillary length $\ell = \gamma/G_0 \ll h_s$, a characteristic time scale $\tau = \eta_s/G_0$, and a characteristic velocity $v_{\ell} = \ell/\tau=\gamma/\eta_s$.

\begin{table}%
    \caption{\label{tab:st:parameters}Material parameters}%
    \begin{center}\begin{tabular}{ccc}
        symbol & value & meaning\\\hline
        $\eta$ & $\unit[1]{mPa\,s}$ & liquid viscosity\\
	    $\gamma$ & $\unit[38]{mN/m}$ & liquid surface tension\\
    	$\gamma_s$ & $\unit[38]{mN/m}$ & solid surface tension\\
	    $G_0$ & $\unit[1]{kPa}$ & static shear modulus\\
	    $\eta_s$ & $\unit[3]{Pa\, s}$ & substrate viscosity\\
    	$h_s$ & $\unit[1]{mm}$ & substrate thickness\\
    	$\epsilon$ & $\unit[4.75]{\upmu m}$ & interface thickness\\\hline
    	$\ell$ & $\unit[38]{\upmu m}$ & elastocapillary length\\
	    $\alpha_s = \frac{\gamma_s}{G_0\,h_s}$ & $0.038$ & elastocapillary number\\
    	$v_{\ell}=\gamma_s/\eta_s$ & $\unit[0.0126]{m/s}$
    	& characteristic velocity
    \end{tabular}\end{center}%
\end{table}

Our numerical model is formulated with cylindrical symmetry, implementing the two-phase fluid by a phase-field approach.
Thus the liquid-ambient interface has a finite thickness $\epsilon \ll \ell \ll h_s$, and the capillary traction of the meniscus onto the solid is distributed over this characteristic width.
The solid is modelled by a finite element approach, with a sharp interface toward the fluid.
The grid size is about $5\%$ of the elastocapillary length at the liquid-ambient interface, and typically about $20\%$ outside of the interface region.
All phases are fully coupled to each other by kinematic and stress boundary conditions.
The material parameters are listed in table~\ref{tab:st:parameters}, details on the numerical model are given in~\cite{AlandMokbel2020} and the supplementary material~\cite{supplementary}.
The liquid meniscus is forced to move by imposing the fluxes  $\Phi$ on either end of the cylinder, but can freely change its shape (curvature) in response to the the fluid flow.
Thus the instantaneous contact line velocity $v$ is not imposed, but rather it's long-term mean $\overline{v} = \Phi/(\pi\,R^2)$. All simulations are started at $t=0$ with a flat substrate, a flat meniscus, and a constant imposed flux at the boundaries, and run until a steady state or limit cycle has been reached.

We compare our simulation results for the solid deformation to the analytical plane-strain model from~\cite{Karpitschka:NC2015}, imposing a constant contact line velocity $v=\overline{v}$ and replacing the $\delta$-shaped traction by
\begin{equation}
    \label{eq:phasetraction}
	T(x) = \frac{3 \gamma}{4 \sqrt{2}\, \epsilon }  \left(1-\tanh ^2\left(\frac{x - v t}{\sqrt{2}\,\epsilon }\right)\right)^2,
\end{equation}
which can be derived for the phase field model in equilibrium (see supplementary material for details~\cite{supplementary}).
Since $h_s\ll R$, the substrate deformation is well approximated by plane-strain conditions. 
Importantly, since $\epsilon\ll\ell$, our analytical and numerical results do not significantly depend on the actual value of $\epsilon$.
Figure~\ref{fig:simulation_profiles}~(b) shows the quasi-stationary substrate deformation for several imposed velocities, comparing simulation results (markers) to the analytical model (lines), in excellent agreement.
Note, that this comparison is only possible as steady ridge shapes are observed for the chosen velocities. In an intermediate velocity range we find unsteady cyclic shape dynamics, as detailed below, which cannot be grasped by the analytical model.

\section{Modes of contact line motion}%
\begin{figure}\begin{center}%
	\includegraphics{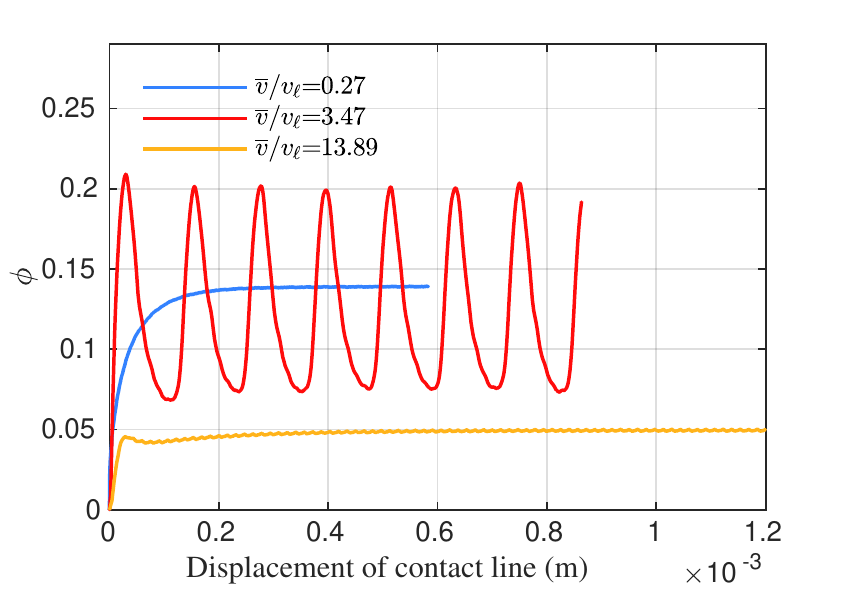} 
	\caption{Rotation $\phi$ of the liquid-ambient interface at the contact line, as a function of the contact line position, for different imposed mean velocities. Slow and fast speeds show a continuous motion (blue and yellow, respectively). For intermediate velocities (red) we observe a strong stick-slip behavior, in which the liquid angle oscillates with an amplitude that is comparable to its mean.}
	\label{fig:simulation_angle_displacement}%
\end{center}\end{figure}%
The dynamics of the contact line motion are characterized by the time-dependent rotation $\phi = \theta-\theta_{eq}$ of the liquid interface at the triple line (cf. Fig.~\ref{fig:simulation_profiles}~(a)).
Figure~\ref{fig:simulation_angle_displacement} shows $\phi$ as a function of contact line position for the three characteristic regimes that we find in our simulations.
At small speed ($v\lesssim v_{\ell}$, blue), after some initial transient the contact line moves steadily, with a constant dynamic contact angle.
Here, the relation between $v$ and $\phi$ is permanently dominated by viscoelastic braking~\cite{Carre:N1996,Long:L1996,Karpitschka:NC2015}.
Once the forcing velocity exceed a critical value, the motion becomes unsteady, finding a limit cycle after an initial transient (red): the liquid interface rotation $\phi$ shows large oscillations, of peak-to-peak amplitude $\Delta\phi$ on the order of the mean rotation $\overline{\phi}$, with a non-trivial waveform, as the contact line advances.
This behavior is not captured by the simple analytical model.
For larger speeds (yellow), we observe again a constant $\phi$ after an initial transient. We note here that the motion in this regime is very sensitive to discretization artifacts and requires rather fine grid resolutions to give consistent results.
Movies illustrating contact line motion and substrate dynamics for the three modes in Figure \ref{fig:simulation_angle_displacement} can be found in the supplementary data.

\begin{figure}\begin{center}%
	\includegraphics{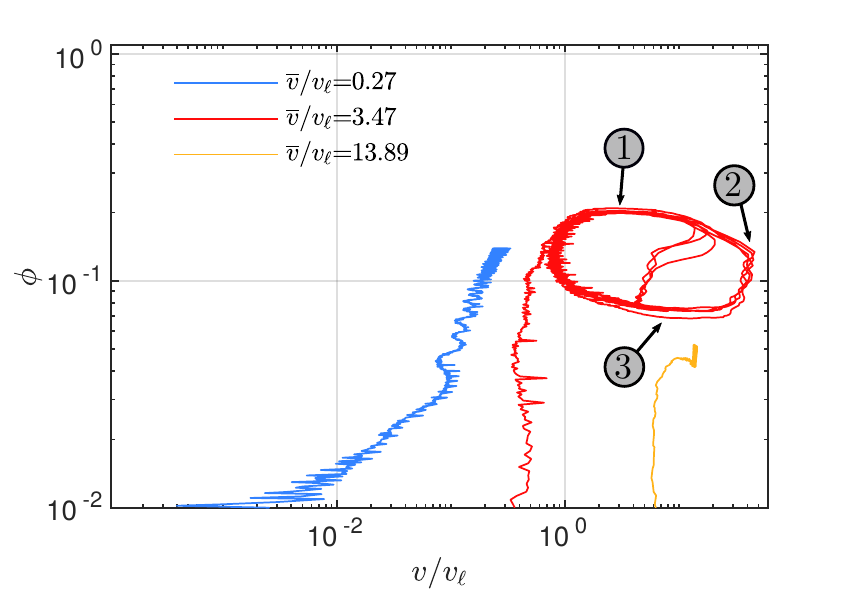}%
 \caption{Phase portraits for contact line motion. Blue: stable, stationary-motion regime. After an initial transient, the contact line finds a stationary constant value in the $v$-$\phi$-plane. Discretization artifacts are visible for small speeds (mind the logarithmic scale scale). Red: stick-slip motion, characterized by a large limit cycle in the $v$-$\phi$-plane. Yellow: at large speeds, stick-slip motion is suppressed, finding stationary point in the $v$-$\phi$-plane again.
 }%
	\label{fig:phaseportrait}%
\end{center}\end{figure}%

Figure~\ref{fig:phaseportrait} shows a phase portrait of the contact line motion i.e., in terms of the physically relevant variables $\phi$ and $v$:
$\phi=\theta-\theta_{eq}$ is a measure of the imbalance in Young's equation, and thus a measure of the total dissipative force (liquid and solid).
Multiplied with the instantaneous velocity, one obtains the total dissipated power per unit length of contact line, since our equations of motion are overdamped.
For slow forcing speeds (blue), we observe a continuous, steady contact line motion, up to the scale of grid artifacts (mind the logarithmic scales).
For intermediate forcing speeds (red), we observe a limit-cycle: As the liquid rotation exceeds a well-defined maximum (\raisebox{.5pt}{\textcircled{\raisebox{-.9pt} {1}} in Fig.~\ref{fig:phaseportrait}}), the contact line accelerates.
In this phase, it surfs down it's own wetting ridge, releasing energy stored in the meniscus curvature, rate-limited partly by liquid dissipation (\raisebox{.5pt}{\textcircled{\raisebox{-.9pt} {2}}} in Fig.~\ref{fig:phaseportrait}).
It thus decelerates, and a new wetting ridge starts to grow, opposing the contact line motion further (\raisebox{.5pt}{\textcircled{\raisebox{-.9pt} {3}}} in Fig.~\ref{fig:phaseportrait}) until the next cycle starts.
For larger forcing speeds, the region covered by the limit cycle decreases until it virtually vanishes (yellow), up to grid artifacts.
This is caused by the growing importance of liquid dissipation, which effectively limits, and finally prevents, the large-speed excursions during the slip phases.

\section{Regimes of contact line motion}%
\begin{figure}\begin{center}%
	\includegraphics{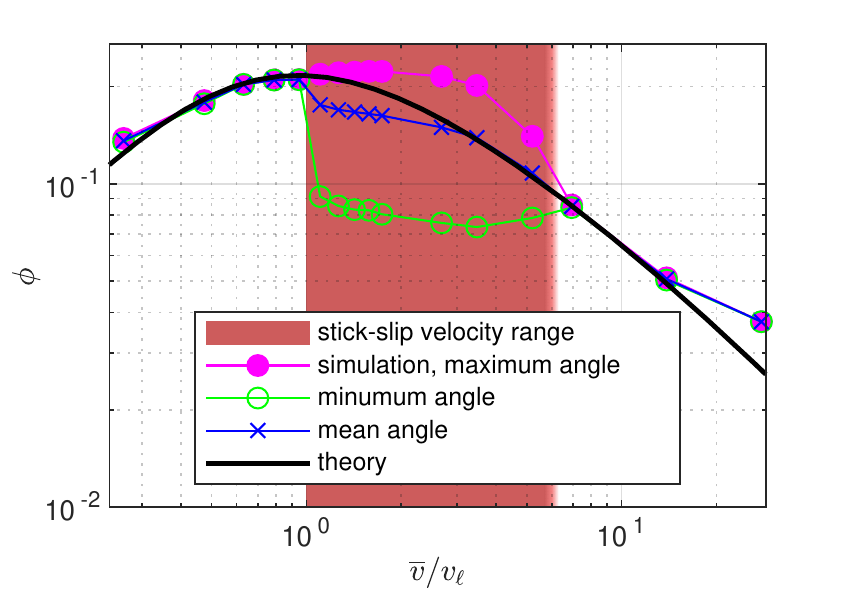}%
 \caption{Rotation $\phi$ of the liquid-ambient interface relative to its equilibrium orientation, as a function of the imposed mean velocity $\overline{v}$. In the red region, the contact line motion is unsteady (stick-slip) in the simulations. The solid black line depicts the analytical calculation of the ridge tip rotation, for an imposed constant contact line velocity. Markers depict the maximum, mean, and minimum angle observed in the simulations. The onset of unsteady motion correlates with the maximum in ridge rotation. At large speeds, the amplitude of the angle oscillations decreases and the motion becomes stationary again until, finally, liquid dissipation becomes relevant.}%
	\label{fig:simulation_angle_v}%
\end{center}\end{figure}%
We characterize the contact line motion by $\phi_{max}$ (\raisebox{.5pt}{\textcircled{\raisebox{-.9pt} {1}}} in Fig.~\ref{fig:phaseportrait}), $\phi_{min}$ (\raisebox{.5pt}{\textcircled{\raisebox{-.9pt} {3}}} in Fig.~\ref{fig:phaseportrait}), and $\overline{\phi}$, the minimum, maximum, and mean values of $\phi$ in the stationary/limit cycle regime. Figure~\ref{fig:simulation_angle_v} shows these values as a function of the imposed (long-term mean) $\overline{v}$.
For small speeds, $v=\overline{v}$, and the simulated $\phi$ (symbols) coincides with the result of the analytical model (black line), indicating the stability of steady contact line motion.

The onset of stick-slip motion (red region) aligns with the maximum of $\phi$ observed in the analytical model where $v$ is imposed instead of $\overline{v}$.
This was stipulated in~\cite{Karpitschka:NC2015} since the rotation $\phi$ is a measure for the dissipative (viscoelastic braking) force: A dissipative force that decreases with speed causes acceleration, and thus an unstable motion.
The maximum braking force is observed at $\overline{v} = v_{\ell} = \gamma/(G_0\,\tau)$, the elastocapillary velocity:
The finite width of the traction distribution regularizes the dissipation singularity at the scale $\epsilon\ll\ell\ll h_s$.
Thus the contact line motion excites a dominant frequency $\sim \overline{v}/\epsilon$ in the solid, corresponding to a dynamical elastocapillary length $\ell_{\overline{v}}\sim\gamma_s/G^*(\overline{v}/\epsilon) = \gamma_s\,\epsilon/(\eta_s\,\overline{v})$.
Resonance is expected at $\epsilon\sim\ell_{\overline{v}}$ i.e., $\overline{v}\sim\gamma_s/\eta_s$, independent of the choice of $\epsilon$, which is confirmed in our analytical model~\cite{supplementary}.
$\phi_{max}$ remains approximately constant upon entering the stick-slip regime, indicating a well-defined upper limit of the viscoelastic braking force also in unsteady situations (cf. location \raisebox{.5pt}{\textcircled{\raisebox{-.9pt} {1}}} in Fig.~\ref{fig:phaseportrait}).
However, this force periodically drops to much smaller values, as indicated by the much smaller values of $\phi_{min}$.
In these \emph{surfing} phases (\raisebox{.5pt}{\textcircled{\raisebox{-.9pt} {2}}} in Fig.~\ref{fig:phaseportrait}), liquid dissipation and the finite capillary energy stored in the curved meniscus are the rate-limiting factors.

As the imposed $\overline{v}$ is increased further, the amplitude of the oscillation $\Delta\phi$ shrinks, reaching virtually zero (indicated by the fading red region). 
In this regime, the reduced viscoelastic braking force ($\phi_{max}$) limits the build-up of capillary energy in the meniscus, while liquid dissipation prevents its fast release.
Thus the oscillatory motion is effectively damped out by liquid dissipation, while the overall motion is still governed by viscoelastic braking:
$\overline{\phi}\sim \overline{v}^{-1}$ closely follows the result from the analytical model.
The increased mean liquid rotation for the largest velocity $\sim 0.28 v_{\ell}$, relative to the prediction of the analytical model, is caused by liquid dissipation. This can be rationalized by a comparison with the Cox-Voinov law for moving contact lines on rigid surfaces, which, given the capillary number $\sim 10^{-2}$, predicts rotations on this order of magnitude~\cite{Cox:JFM1986,Voinov:FD1977,Snoeijer:ARFM2013}.
In this hydrodynamic regime, one returns to the classical wetting physics on rigid surfaces.

\begin{figure}\begin{center}%
    \includegraphics{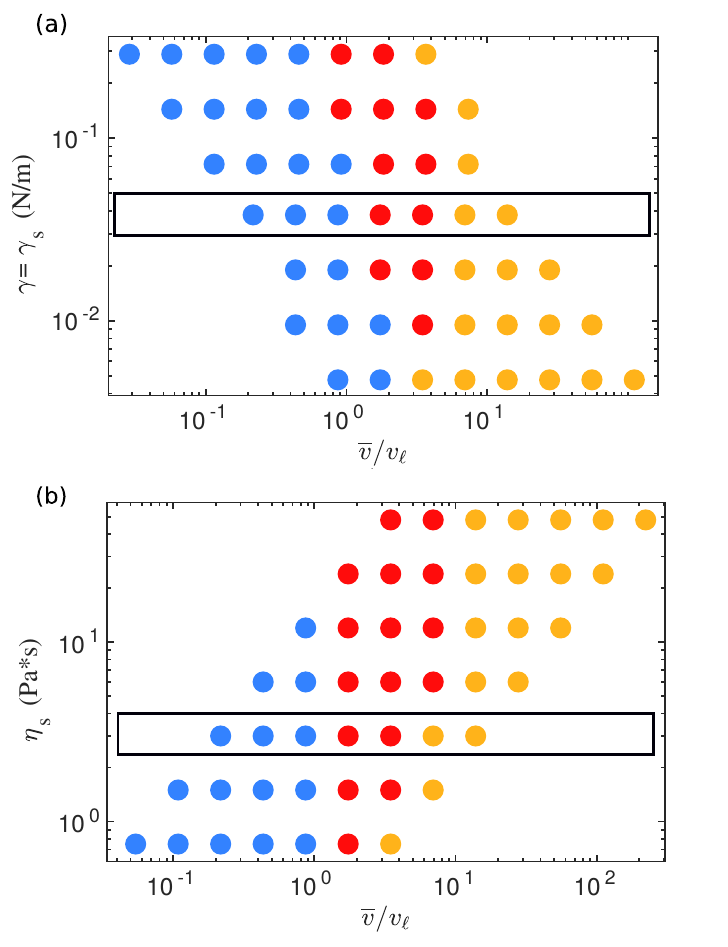}
 \caption{Phase diagrams for stick-slip behavior vs. contact line speed. Blue: steady contact line motion; red: stick-slip; yellow: high-speed continuous motion. (a) Tuning the magnitude of capillary forces $\gamma = \gamma_s$ on the vertical axis. (b) Varying substrate viscosity on the vertical axis.}%
	\label{fig:simulation_Phasediagram}%
\end{center}\end{figure}%

In Figure~\ref{fig:simulation_Phasediagram}, we summarize the dynamical wetting behavior in terms of these three modes, as a function of the imposed mean speed and the solid parameters.
Steady small-speed, stick-slip, and steady high-speed modes are indicated by blue, red, and yellow discs, respectively.
On panel (a), we vary on the vertical axis the solid and liquid surface tensions and thus the elastocapillary number $\alpha_s = \ell / h_s$, while keeping the Neumann angles of static wetting constant.
The onset of stick-slip is located near $\overline{v}=v_{\ell}$, given by the maximum of $\phi$ vs. $v$. This maximum is independent of $\alpha_s$, up to a small correction due to the finite thickness $\epsilon$ of the fluid interface, as can be shown by the analytical model (see Fig.~1 of the supplementary material~\cite{supplementary}).
In physical units, however, the onset of stick-slip is inversely proportional to $\gamma_s$ since $v_{\ell} \sim \gamma_s$.
The transition to the fast continuous mode is, in scaled units, nearly independent of the surface tension.\
Consequentially, the stick-slip mode disappears at very low $\gamma$.

Similarly, the solid viscosity $\eta_s$ (panel (b)) has no measurable impact on the critical $v/v_{\ell}$ for the transition to stick slip, but the physical critical velocity is proportional to $\eta_s$ since $v_{\ell}\sim\eta_s^{-1}$.
Thus, for small solid viscosities, the damping effect of liquid dissipation and thus the transition back to steady motion becomes noticeable already at smaller $\overline{v}/v_{\ell}$, such that the stick-slip region ultimately disappears at very low $\eta_s$.
In any case, at very large speeds, liquid dissipation will take over, leading to  wetting dynamics equivalent to those on rigid surfaces.

\section{Discussion}%
In this Letter, we provide a comprehensive numerical analysis of dynamical soft wetting, including the physics of all relevant elements, the liquid, the solid, and the interfaces.
For each element, we used the minimal required level of complexity, to keep physics intact and conceivable: The Stokes limit for the fluid, a Kelvin-Voigt constitutive relation for the soft solid, regularized at a constant scale $\epsilon\ll\ell$, and constant and equal solid surface tensions on all three interfaces.
This simple model already requires a complex, strongly coupled multi-physics modelling approach, and exhibits rich behaviors.

Our numerical experiments cover a wide range of system parameters, and we reveal three regimes in which the dominant physical mechanisms differ:
(i) a slow regime, in which the contact line motion is entirely dominated by the the dissipation in the solid.
This regime is observed as long as the viscoelastic braking force increases with speed.
(ii) an intermediate regime, in which the dominant rate-limiting mechanism periodically switches from solid to liquid dissipation.
This regime starts where the viscoelastic braking force exhibits a maximum with respect to the contact line velocity. This maximum is caused by a resonance effect, due to the regularization of a singular dissipation at some finite (constant) length scale.
Other mechanisms, like dynamic solid surface tensions (surface constitutive relations~\cite{Xu:SM2018,Xu:NC2017,Gorcum:PRL2018,Liu:SM2020,Heyden:PRSA2021,Zhao:SM2021}) or a constitutive relation that exhibits resonance (e.g., standard-linear solid~\cite{Karpitschka:NC2015}) would lead to the same phenomenology.
(iii) a large-$\overline{v}$-regime with continuous motion, yet governed by viscoelastic braking, in which liquid dissipation prevents strong oscillations of the meniscus.
Since the viscoelastic braking force, in contrast to liquid dissipation, does not increase with velocity, we ultimately find again liquid dissipation dominating the contact line motion, and one recovers the wetting physics of rigid surfaces.

With this first comprehensive overview of soft wetting physics scenarios, we provide a strong basis for interpreting different phenomenology observed in experiments, ranging from paraffins~\cite{Kajiya:SM2013} over microelectronic sealants~\cite{Gorcum:PRL2018,Style:PRL2013} to biology~\cite{Holly:EER1971,Prakash:S2008,PerezGonzalez:NP2018}, and motivate experiments in the so-far little explored large-$\overline{v}$ regimes.

\acknowledgments{SK and SA acknowledge funding by the German Research Foundation (DFG project no. KA4747/2-1 to SK and AL1705/5-1 to SA).  Simulations were performed at the Center for Information Services and High Performance Computing (ZIH) at TU Dresden.}


\end{document}